# Weighted Histogram Equalization Using Entropy of Probability Density Function

# (The Best Paper Award)

Thaweesak Trongtirakul [1] and Sos S. Agaian[2]

[1]Department of Electrical Engineering, Faculty of Industrial Education,
Rajamangala University of Technology Phra Nakhon, Thailand, thaweesak.tr@rmutp.ac.th,

[2]Graduate Center, City University of New York, United States of America, sos.agaian@csi.cuny.edu

**Abstract**

Low-contrast image enhancement is essential for high-quality image display and other visual applications. However, it is a challenging task as the enhancement is expected to increase the visibility of an image while maintaining its naturalness. In this paper, the weighted histogram equalization using the entropy of the probability density function is proposed. The computation of the local mapping functions utilizes the relationship between non-height bin and height bin distributions. Finally, the complete tone mapping function is produced by concatenating local mapping functions. Computer simulation results on the CSIQ dataset demonstrate that the proposed method produces images with higher visibility and visual quality, which outperforms traditional and recently proposed contrast enhancement algorithms methods in qualitative and quantitative metrics.

## 1. Introduction

Low contrast images have a detrimental effect on visualization. Contrast enhancement focuses on eliminating the problem of highly concentrated contrast in the specific segments of the whole dynamic range. Several approaches are reviewed in the literature to amend the low contrast visualization issue. Histogram Equalization (HE)-based methods [1] are commonly used for contrast enhancement. The traditional HE algorithm [1] is the low-cost computational, straightforward, and effective contrast enhancement method. The results are not optimum as the enhanced images contain artifacts, and the global brightness remains stable in the average luminance level.

Numerous variants are suggested in the HE process to surmount the general limitations. Contrast Limited Adaptive HE (CLAHE) [2] is used to improve the local contrast. To conserve the mean illumination of the image, much research has introduced a sub-histogram equalization technique [3-6] by separating the histogram into several sub-segments, and HE applies to the individual sub-segments. These modified versions of HE successfully preserve mean brightness. Occasionally, these algorithms are not suitable for various levels of distortion. Recently, Srinivas *et al*. propose a context-based image contrast enhancement using energy equalization with clipping limit, in which the energy curve utilizes a modified Hopfield Neural Network (HNN) architecture [7]. It attempts to yield optimum contrast enhancement with the preserved natural characteristics of the original image. However, the algorithm requires a large number of datasets to learning in the modified HNN architecture. In the case of an ultra-low contrast appearance, the HNN cannot provide the proper suitable set of parameters to the enhancement mechanism.

Another class of image enhancement, the Retinex image representation, carries out a solid and flexible framework for low-light image enhancement. In the Retinex theory, a given image, $I_{i,j,k}$, can be represented as the product of an illumination component, $\mathcal{L}_{i,j,k}$, and a reflectance component, $\mathcal{R}_{i,j,k}$. In general, the Retinex-based methods suffer from solving the ill-posed $I_{i,j,k}$ and $\mathcal{R}_{i,j,k}$ decomposition. With $I_{i,j,k}$ and $\mathcal{R}_{i,j,k}$, [8-10] the low-light and low-contrast enhancement can be easily realized as $I_{i,j,k}^{\gamma} \cdot R_{i,j,k}$, where $I_{i,j,k}^{\gamma}$ denotes the Gamma Correction (GC) that non-linearly transforms the illumination distribution ($\gamma$ is empirically set as $0.\dot{4}\dot{5}$). However, limitations still exist in these Retinex-based methods. Hao *et al*. address these issues by proposing a simple semi-decoupled Retinex decomposition [11], in which the Retinex image decomposition is achieved in an efficient semi-decoupled. The method illustrates good performances in suppressing noise and retains the natural visual appearance. However, the technique fails to demonstrate an informative image when the original image contains extreme low-light phenomena with a low-contrast appearance.

In this paper, the main focus is introducing a new weighted histogram equalization based on the entropy of the probability density function (PDF). The calculation of distribution weights using the entropy of PDF is used for the image enhancement task. The computed weights relate to the relationship between height bin and non-height bin distributions. The Entropy Measure of Enhancement (EME) [12-14] is applied to generate the best contrast-enhanced image during contrast optimization.

The remaining sections of the paper are structured as follows. In section II, the weighted histogram equalization with the entropy of the PDF is presented. Section III exhibits the results and discussion of the proposed method. Finally, the paper ends with the conclusion in Section IV.

## 2. Weighted Histogram Equalization with the Entropy of PDF

The proposed method is illustrated graphically in Fig. 1. It comprises two major steps: PDF entropy-based threshold computation and visibility optimization. The PDF entropy-based threshold computation and optimization process are described in the following sub-sections.

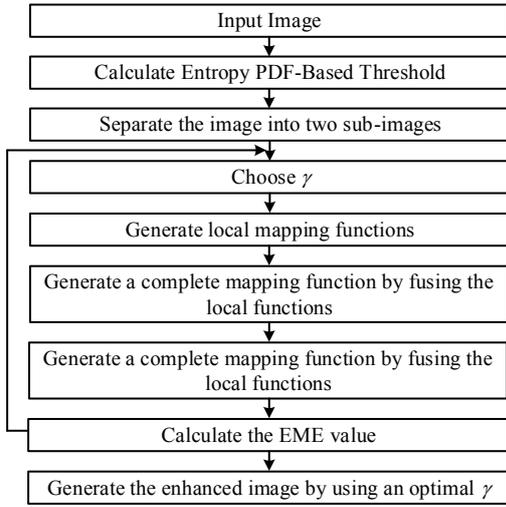

Fig. 1. Weighted histogram equalization with the entropy of pdf framework

*Entropy PDF-Based Threshold:* For an image, the threshold of the entropy PDF is derived from the fundamental of the Entropy Measure of Enhancement (EME) architecture [12-14]. The EME architecture comprises the entropy of the contrast ratio between lower and upper bounds. In our implementation, the threshold of an image can be described as:

$$t = arg\max_x \left(\frac{1}{x_{L-1}}\sum_{x=1}^{x_{L-1}} e(x)\right) \quad (1)$$

$$e(x) = \frac{\sum_{x=x_t+1}^{x_{L-1}} p_u(x) - \sum_{x=1}^{x_t} p_l(x) + \alpha}{\sum_{x=x_t+1}^{x_{L-1}} p_u(x) + \sum_{x=1}^{x_t} p_l(x) + \alpha} \cdot \log\left(\frac{\sum_{x=x_t+1}^{x_{L-1}} p_u(x) - \sum_{x=1}^{x_t} p_l(x) + \alpha}{\sum_{x=x_t+1}^{x_{L-1}} p_u(x) + \sum_{x=1}^{x_t} p_l(x) + \alpha} + \beta\right) \quad (2)$$

where $p_l$ and $p_u$ respectively denote the lower and the upper PDF subsets separated by a threshold, $x_t$. $\alpha$ and $\beta$ are a constant. $x_{L-1}$ represents the total number of luminance levels.

*Optimal Weighted Histogram Equalization:* in this sub-section, generated tone mapping functions are calculated independently and combined with mapping the complete mapping function. The computation of the tone mapping function is similar to the Histogram Equalization (HE) process. It involves three steps: calculation of PDF, calculation of CDF, and mapping function generation. The PDF for sub-images denotes $\boldsymbol{p_l}(x)$, and $\boldsymbol{p_u}(x)$ defined as

$$\boldsymbol{p_l}(x) = \frac{card(I_{i,j}(x))}{N} \ ; \ x \leq t \quad (3)$$

$$\boldsymbol{p_u}(x) = \frac{card(I_{i,j}(x))}{N} \ ; \ x > t \quad (4)$$

where $I_{i,j}$ denotes a given image, which contains grayscale levels, $x$. $card(\cdot)$ represents the cardinality operation. $N$ is the total number of pixels in an image. $t$ refers to the optimal threshold, which can be calculated by Eq. (1). The CDF for each sub-image is computed as

$$c_l(x) = \sum_{a=1}^{b} \boldsymbol{p_l}(a) \ ; \ 1 \leq b \leq t \quad (5)$$

$$c_u(x) = \sum_{a=1}^{b} \boldsymbol{p_u}(a) \ ; \ t+1 \leq b \leq x_{L-1} \quad (6)$$

The tone mapping function for each sub-image is generated as

$$T_l(x,\gamma) = x_0(1 - \omega_l(x)^\gamma) + (t - x_0)c_l(x)\omega_l(x)^\gamma \quad (7)$$

$$T_u(x,\gamma) = t + 1 + (1 - \omega_u(x)^\gamma) + (x_{L-1} - t)c_u(x)\omega_u(x)^\gamma \quad (8)$$

The proposed method transforms the intensity distribution like the HE algorithm using the transformation function derived from Eq. (9), where $\omega_l(x)$ and $\omega_u(x)$ refer to the distribution weights, $\omega_l(x) = 1 - \frac{k_l}{t}, \omega_u(x) = 1 - \frac{k_u}{x_L - t}$, which denote the non-height bin-to-height bin ratio as illustrated in Fig. 2. The complete mapping function is generated by concatenating the tone mapping function for each sub-image as

$$T(x,\gamma) = T_l(x,\gamma) \cup T_u(x,\gamma) \quad (9)$$

Where ∪ denotes a concatenating operation. The resulting contrast-enhanced image is formulated by applying Eq. (9) to the given image.

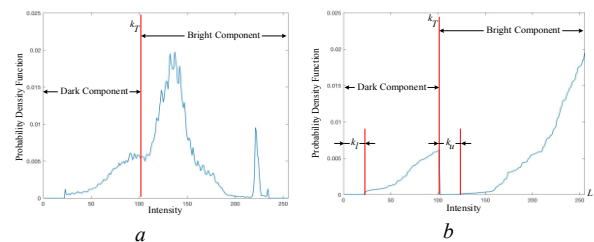

a) Original image histogram with separation point $k_T$, dark component, and bright component sub-histograms. b) Unionization of Modified PDFs from Fig. 2(a)

Fig. 2. Histogram separation and intensities arrangement

The given intensity distribution to the output intensity values is based on the transformation function. It may be visualized as a lookup table, where each luminance level in the given image is transformed to the output luminance value in the enhanced image

$\boldsymbol{\gamma}$-*Optimal Estimation*: in the proposed histogram-based equalization algorithm, the power factor, $\boldsymbol{\gamma}$, plays a significant role in distributing luminance levels. The power factor can be optimized by using a greedy optimization as

$$\gamma_\Omega = \arg\max_\gamma \left\{ EME\left(T(I_{i,j,k}, \gamma)\right)\right\} \quad (10)$$

$$EME = \frac{1}{M \times N} \sum_{i=1}^{M} \sum_{j=1}^{N} \left(f(I_{i,j,k})\right) \log\left(b + f(I_{i,j,k})\right) \quad (11)$$

$$f(I_{i,j,k}) = \left( \frac{[I_{max}]_{i,j,k}^{m,n,3} - [I_{min}]_{i,j,k}^{m,n,3} + a}{[I_{max}]_{i,j,k}^{m,n,3} + [I_{min}]_{i,j,k}^{m,n,3} + a} \right) \quad (12)$$

where $\gamma$ denotes the set of a power factor, $\gamma = \{0.1, 0.2, \ldots, 1.0\}$. $[I_{min}]_{i,j}^{m,n}$ and $[I_{max}]_{i,j}^{m,n}$ respectively refer to the local minimum and the local maximum luminance levels in a considered block, $m \times n$. To avoid logarithmic errors and negative values, $a$ and $b$ represent a constant, $a = 1, b = 1$.

## 3. Results and Discussions

In our experiments, the proposed algorithm runs on the HSV color-space. The S and V are considered. Various experiments are conducted to demonstrate the proposed algorithm. The experiments are engaged with images of different illumination conditions with significant contrast variations. Since this is the optimally weighted histogram equalization with the entropy of PDF is utilized for image enhancement, the performance is compared with the other state-of-the-art histogram-based image enhancement algorithms. Therefore, the comparison is justified.

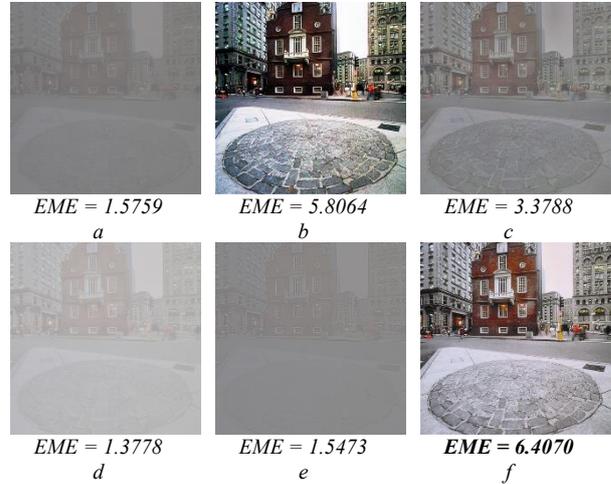

a) original image. b) HE [1]. c) CLAHE [2]. d) LIME [7]. e) LR3M [11]. and f) proposed method.

Fig. 3 Visual representation of input and enhanced images (CSIQ [15])

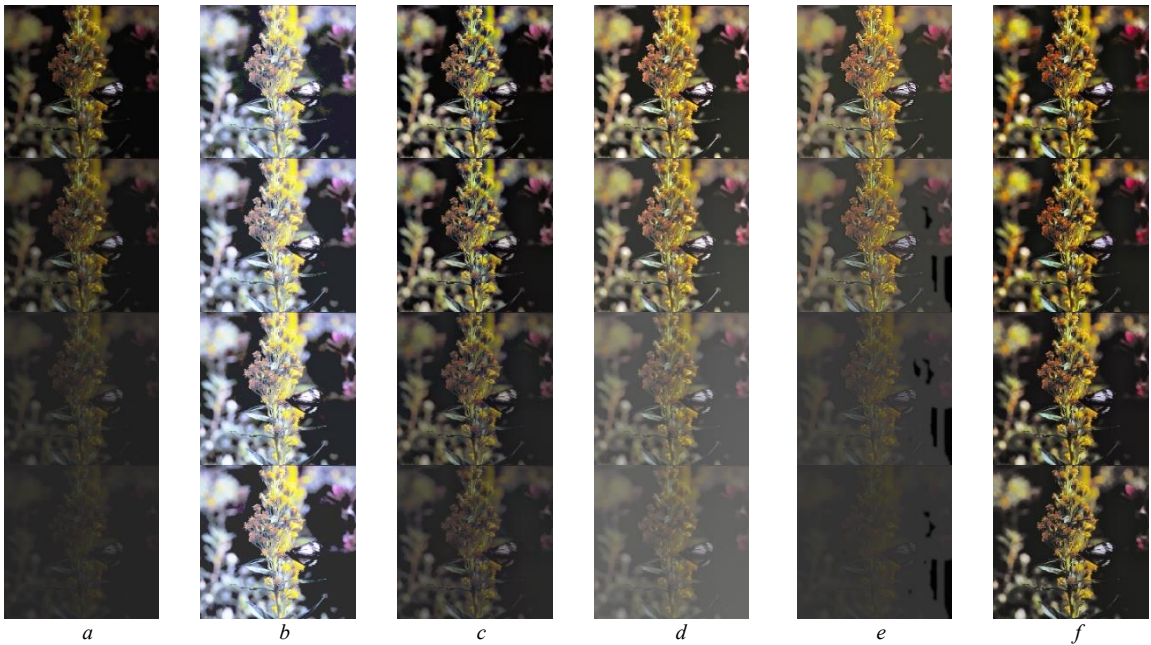

a) original image. b) HE [1]. c) CLAHE [2]. d) LIME [7]. e) LR3M [11]. and f) proposed method.

Fig. 4 Visual representation with different distortions of input and enhanced images (CSIQ [15])

*Qualitative Assessment*: The proposed algorithm has been applied to several hundred images from the CSIQ database provided by the Oklahoma State University [15]. The database consists of 30 original images; each is distorted using six different types of distortions at four to five different distortion levels. The resulting enhanced images with the proposed algorithm have been compared with commonly used histogram-based image enhancement algorithms and existing image enhancement algorithms. These methods include HE [1], CLAHE [2], LIME [7], and LR3M [11]. The visual comparison is presented in Figs. 3-4. The input image has a low-contrast appearance, whereas the compared methods produce unpleasant visualization in the enhanced images. For the HE image, some regions suffer from an under-enhancement artifact. The CLAHE, LIME, and LR3M fail to illustrate distinctive and informative visualization. In the case of the proposed method, it produces an optimal enhancement level without introducing unpleasant artilects.

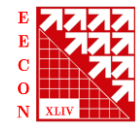

Fig. 4 represents the input and enhanced image with different distortions. This result shows that the HE method yields over-enhancement and brightness saturation. The CLAHE [2], LIME [7], and LR3M [11] introduce a similar effect: the differences of varying ground objects are still tiny. Compared with other methods, the proposed method yields a significant improvement in spatial details and color appearances, demonstrating that our algorithm results with more clarity and the different ground objects in the enhanced image are easily distinguishable.

*Quantitative Assessment:* Human Visual System (HSV) is used to judge the image enhancement quality. Image information or visual image quality is highly related to how humans perceive that may vary from human to human. The image enhancement procedure may also illustrate undesirable artifacts. Hence, in encouraging the qualitative assessment, a quantitative assessment is also mandatory. Therefore, image quality assessment using quality metrics has been available. A faith contrast enhancement algorithm ought to obtain an adequate enhancement level with maximum contrast. The Entropy Measure of Enhancement (EME) [12-14] is used to justify image enhancement quality.

In addition, we compared our method with classical and current state-of-the-art methods [1-2,7,11]. The results are demonstrated in Table 1. It shows that the proposed method outperforms other methods and gets competitive results with the highest score according to these no-reference image quality assessments. The EME confirms that the proposed method achieves to improve the visibility of the scene restoration.

Table 1 Quantitative Assessment of Figure 4 by Using EME

| Methods | Distortion Levels | | | |
|---|---|---|---|---|
| | - | Light | Moderate | Heavy |
| Original | 5.7736 | 4.1259 | 2.0459 | 1.3893 |
| HE | 4.9540 | 4.0323 | 3.9904 | 3.4375 |
| CLAHE | 5.9717 | 5.0946 | 3.6830 | 2.5996 |
| LIME | 5.1021 | 3.6968 | 1.7462 | 1.1285 |
| LR3M | 4.7971 | 3.7018 | 1.8681 | 1.2802 |
| Proposed | **7.1244** | **6.1243** | **4.8905** | **4.1960** |

## 4. Conclusion

In this paper, a new weighted histogram equalization using the entropy of the probability density function has been introduced. The image histogram is petitioned into two sub-images based on the entropy of the probability density function. The tone mapping functions include the relationship of luminance distribution. The computed weight is chosen as the maximum contrast enhancement score (EME). Each sub-image is equalized independently and fused to form the complete enhanced image. The proposed method restores scene visualization while avoiding under-enhancement and over-enhancement artifacts. Also, the submitted images illustrate a natural-looking appearance. The proposed method has been extensively validated through qualitative and quantitative experiments, revealing that it competes well and outperforms both classical and current-state-of-the art methods. The method has been shown to be robust, efficient, and capable of correctly removing the visual effect of contrast on images acquired under a variety of contrast degradation.

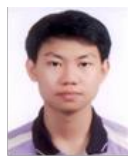

**Thaweesak Trongtirakul** received the D.Eng. degree in Electronics and Telecommunication from King Mongkut's University of Technology Thonburi, Bangkok, Thailand. His primary research interests are in computer vision, machine vision, optical images, and smart cities.

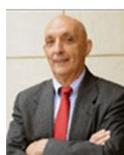

**Sos Agaian** (M'98–SM'00) received the M.S. degree (summa cum laude) in mathematics and mechanics from Yerevan University, Yerevan, Armenia, the Ph.D. degree in math and physics from the Steklov Institute of Mathematics, Russian Academy of Sciences, Moscow, and the Doctor of Engineering Sciences degree from the Institute of the Control System, Russian Academy of Sciences. He is a Distinguished Professor of Computer Science at the College of Staten Island and the Graduate Center (CUNY).